\documentclass[12pt]{article}

\usepackage[utf8x]{inputenc}
\usepackage{amsthm}
\usepackage{amsmath}
\usepackage{amssymb}
\usepackage{bm}
\usepackage{float}
\usepackage{afterpage}
\usepackage{tabularx}

\usepackage[top=1.7cm, left=1.2cm, right=1.3cm, bottom=2cm]{geometry}

\usepackage{graphicx,xcolor}

\usepackage{xcolor}

\title{Networks with Correlated Edge Processes}
\author{Maria S\"uveges \& Sofia C. Olhede}
\date{\today}

\begin{document}

\maketitle

\begin{abstract}
    This article proposes methods to model nonstationary temporal graph processes. This corresponds to modelling the observation of edge variables (relationships between objects) indicating interactions between pairs of nodes (or objects) exhibiting dependence (correlation) and evolution in time over interactions. This article thus blends (integer) time series models with flexible static network models to produce models of temporal graph data, and statistical fitting procedures for time-varying interaction data. We illustrate the power of our proposed fitting method by analysing a hospital contact network, and this shows the high dimensional data challenge of modelling and inferring correlation between a large number of variables. 
\end{abstract}


\section{Introduction}\label{sec:intro}
This paper introduces time series models for observations of dynamic graphs over time, and methods to estimate these models. This set of developments is motivated by the increasing prevalence of temporal observations of interactions between 
entities in many application areas~\cite{ahmed2009recovering,ribeiro2013quantifying,liu2013time}. We call such observations  {\em dynamic graphs}, and the observations correspond to samples from a temporal graph process~\cite{crane2016dynamic}, rather like a time series can be viewed as samples of a continuous time stochastic process. The aim of this paper is to introduce a natural generalized linear modelling framework for discretely regularly sampled temporal graph processes that can flexibly capture data features such as cyclostationary and dependence of edges in time. 

The rising ubiquity of dynamic graphs has been matched by technical innovation for their analysis, see for example ~\cite{matias2018asemiparametric,ludkin2018dynamic,pensky2019dynamic,jiang2020autoregressive,pamfil2020inference}.  
Simultaneously, the realisation that networks should be described directly in terms of observed interactions or edges rather than in terms of describing the interactions between nodes, in a nodal view, has been gaining considerable traction~\cite{crane2018edge}. These theoretical  developments are paralleled by the recognition that nodal clustering may not be sufficient to model a graph due to overlapping node communities, and this problem may be resolved by assuming the links themselves to form communities on their own. Papers dealing with the detection of link communities and characterising their dynamical behaviour are, for example, \cite{ahn2010link, evans2010clique, kim2011map, nguyen2011adaptive, meng2016incremental}. 

Key to understanding dynamic graphs is proposing models for their dependence and evolution. The basic building blocks must consider the natural invariances of entities and temporal processes, permutation invariance of measure, and shift invariance of measure for stochastic processes. In addition, for non-Euclidean observations such as graphs it no longer makes sense to put all dependence in a zero-mean perturbation, as is done for most time series problems. First, we still want to encode additional temporal structure. We do not want to pose models whose structure is constant over time, and the perturbations are purely random. Second, the temporal structure should be parameterizable, and estimable from one process' realization. Our choice of model will be motivated by a real data example. For this reason we wish to consider models satisfying some permutation invariance at fixed time stamps, but that are not stationary in time. This will be necessary especially as the processes we study could at most be assumed to be cyclostationary.

\begin{figure}[H]
    \centering
    \includegraphics[width=0.98\textwidth]{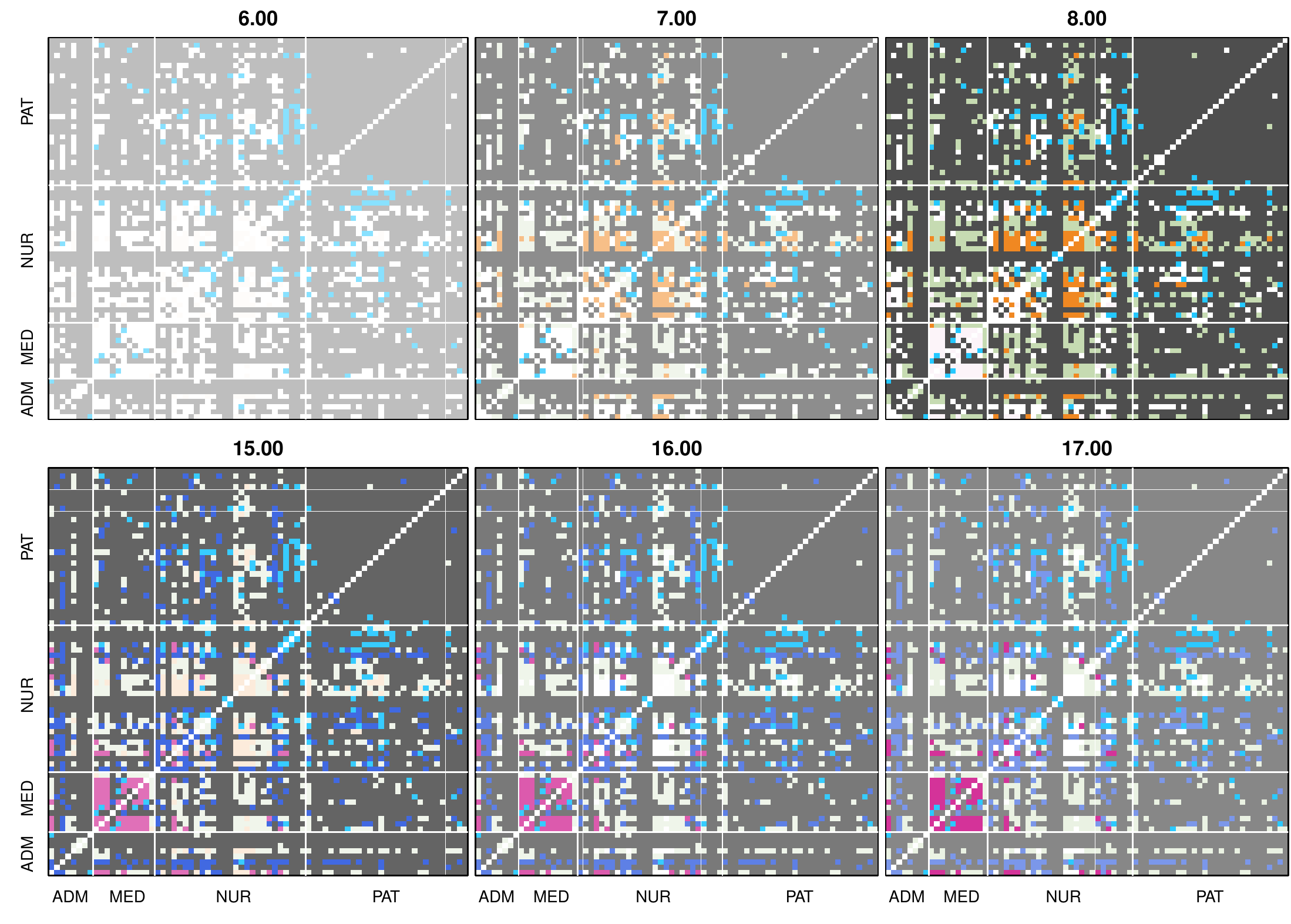}
    \caption{Snapshots of the varying daily activity levels of different link community groups in a hospital ward (see Section \ref{subsec:data_analysis}). The four groups of nodes according to status in the hospital is ADM (administrator), MED (medical staff), NUR (nursing staff) and PAT (patients). The hours of the day are shown in the main title of the panels. The colours indicate the membership of the links, while the intensity of the colours is proportional to the probability of the link being ``switched-on" at that particular time in the day. The maximal intensity of the colour corresponds to the maximal probability of the link to be ``on", which is different for all link communities (they are 0.063, 0.0004, 0.020, 0.011, 0.008 and 0.058 for the orange, black, green, dark blue, light blue, and purple clusters, respectively). White indicates 0 everywhere. }
    \label{fig:linkprobs_vs_t}
\end{figure}

To explain about our modelling framework, and to be sufficiently concrete, let us study an example of a dynamic graph, depicted in
Fig.~\ref{fig:linkprobs_vs_t} as a set of six link communities. It shows the interactions over time of various groups of personnel and patients in a hospital ward in Lyon, France (the data set, the model and the fit resulting in the clustering shown in the Figure is described in Section~\ref{subsec:data_analysis}). As in many other social networks, people in a hospital have distinct and varying contact patterns over time, dictated by a common rhythm of work meetings, patient visits, medical examinations (temporally scheduled 'rounds'), care for patients, meals (also periodic) and so on. The different colours indicate communities of similarly behaving link probabilities over time. The panels show the instantaneous contact probabilities of the communities at different times during the day as intensity of colour, so that for example, deep purple correspond to maximal contact probability of the ``purple" community, and white, to a near-zero contact probability. The upper row shows the morning hours from 6am to 8am, the lower row, the afternoon from 3pm to 5pm. We see that different link communities have very different contact probability at different times of the day. Whereas, for example, the ``green" and ``orange" edges are mostly switched on in the morning, the ``purple" edges activate preferably in the afternoon. 

These empirical facts must be linked with our choice of graph process modelling framework. Our observations are that: first, it is obvious that a model that intends to gain a detailed insight into the dynamics of such a network must be able to account for (periodic/cyclical) time-varying contact patterns, ubiquitous in human contacts. Second, links in human society can show interesting clustering according to their time variation patterns, which can be quite different from a node clustering scheme. Third, it is also evident that temporal evolution and community structure of this network cannot be written in a separable form $\rho(t) f(\xi_i(t),\xi_j(t))$, where $f(\xi_i(t),\xi_j(t))$ is a fixed constant baseline probability of interaction between nodes $i$ and $j$ depending on a latent process $\xi_i$, and $\rho(t)$ is a factor driving the common temporal variation of these probabilities (thus the type of interactions change over time, not just the density of them). Fourth, it also seems plausible that human interactions, especially when observed with high temporal resolution, are in general temporally correlated as interactions cannot come and go willy-nilly from one moment to the other.

It is important to model these patterns, as an alternative to a classical stochastic blockmodel independently generated at each time-point. There are situations where the temporal dynamics of the links is an important factor in the scientific question. An example can be the modelling of the spread of an infectious disease in an evolving community. In this case, a detailed model of the temporal patterns of the contacts may inform much better the public health policymakers about the intervention with optimal cost-efficiency ratio than a descriptive model offering node centrality measures and average contact probabilities, or an SBM putting emphasis on similarities between nodes, not link communities discriminated based on their different temporal dynamics. Stationary processes cannot reproduce all manner of cyclostationary processes common in observations of human activities. Other, similar examples with clear cyclostationary patterns will be found as energy networks or mobile phone networks. The application of detailed dynamic link community models can give a deeper insight as to risks of system breakdowns or overloads.

Currently, there are many proposed methods to perform inference on dynamic networks, say for example~\cite{matias2017statistical, matias2018asemiparametric, ludkin2018dynamic, pensky2019dynamic, jiang2020autoregressive, olivella2021adynamic}, each method coming with either explicit or implicit modelling assumptions. Most of these approaches use the stochastic blockmodel imposing clustering structure on the nodes (SBM; \cite{holland1983stochastic,faust1992blockmodels, anderson1992building, snijders2005estimation}) to model the underlying dynamic structure of the network, with clear choices on how parameters change or evolve across time. 
In the framework of the SBM, dynamics may arise from a latent process describing the evolution of the node memberships in the clusters over time, such as, for example, a set of independent discrete-state Markov processes for each node (e.g., \cite{pensky2019dynamic}) or a hierarchical model allowing for mixed memberships of the nodes and specifying a latent process on the membership distributions (e.g., \cite{olivella2021adynamic}). In most of these models edges are generated independently, perhaps  conditionally on the block memberships of the endpoints of the edges. However, we have seen that many dynamic network observations, such as our example, cannot be taken as a series of temporally independent snapshots, especially in the context of human activities. . 
Exceptions to the description applied in those SBM-based papers cited above are \cite{jiang2020autoregressive} and \cite{ludkin2018dynamic}, where the former constructs correlated graphs by adding correlated noise that may erase or construct edges, that is, introduces correlation at the observed process. The latter models community membership by a switching process, and directly imposes correlation on the edge variables. Thus over a given time interval, correlation between edges is produced.

In this paper we shall model correlation explicitly in the observed edges across time, based on using popular Bernoulli time series models. For the formulation, we reach back to generalised linear models, using their well-known inferential characteristics, and note that methods could be extended to Poisson time series for counts of interactions, see e.g.  \cite{hoff2004modeling, minhas2016new, donnet2019bayesian, schein2014inferring}. In terms of the latent edge variables defining the edge cluster memberships, we shall assume them fixed across time, drawing them at the temporal starting point of the process (assigning a community to each edge). Conditional on link community membership, for each edge we then use the ALARM (A Logistic Autoregressive Model) generating mechanism \cite{agaskar2013alarm}, mainly because this allows the generation of positively and negatively correlated processes within the same model, as explained in Subsection~\ref{subsec:alarmll}.  We then put the ALARM specification in the block model framework, introducing the block-ALARM specification (BALARM model), and give its likelihood, as described in Subsection \ref{subsec:balarmll}. 

We use the EM algorithm to estimate the BALARM model. Whilst a simulation study lets us study the performance under correct model specification, see section~\ref{subsec:simulations}, we also study the performance of this dynamic graph model when analysing temporal social interaction data from the geriatric short-stay ward of a university hospital in Lyon, France~\cite{vanhems2013hospitaldata}, mentioned already above. Using the BALARM model, we uncover groups of interactions between patients and staff, with a clear daily temporal rhythm. This allows us to describe the graph process of temporally correlated edges in a compressed and simple, yet realistic manner, gaining an unprecedentedly detailed insight into the daily activities of a human community.

\section{The Stochastic Blockmodel and the Link Community Model}\label{sec:link}

We start by  recalling the definition of the stochastic blockmodel~\cite{holland1983stochastic,kolaczyk2014statistical}. Consider a network with a fixed set of $N$ nodes, without loss of generality labelled by $\{i : 1, \ldots, N \}$. Let $A_{ij} \in \{0,1\}$ denote the state of the link (edge) between nodes $i$ and $j$, the value 0 indicating the ``switched-off" state of the edge, and 1, its ``switched-on" state.

The matrix $\{A_{ij}\}_{i,j \in \{1, \ldots, N\} }$
is called the adjacency matrix. We model $A_{ij}$ as a Bernoulli variable:
\[
A_{ij} \mid p_{ij} \sim \mathrm{Bernoulli}(p_{ij}),
\]
where $p_{ij} \in [0,1]$ is the probability of link $(ij)$ for being ``on". 

The stochastic blockmodel assumes that each node belongs to one of $K$ possible clusters, and the probability $p_{ij}$ of an edge being ``switched on" is determined by the $K\times K$ matrix $\theta_{ab}$ called the blockmatrix: if node $i$ belongs to cluster $a$ and node $j$ to cluster $b$, then the probability of a link forming between them is equal to $\theta_{ab}$. Introducing the random variable $Z_i \in \{1, \ldots, K\}$ to indicate the cluster membership of node $i$, we can then formulate the stochastic blockmodel as
\begin{align}
   A_{ij} \mid Z_i = a, Z_j = b \,  &\stackrel{\mathrm{iid}}{\sim} \, \mathrm{Bernoulli}(\theta_{ab}),  \\
   Z_i  \, &\stackrel{\mathrm{iid}}{\sim} \, \mathrm{Multinom(\pi_1, \ldots, \pi_K)},
\end{align}
where $\pi_a$ is the probability that a node belongs to cluster $a$. For a dynamical case where we observe the network over time, we need to also specify the temporal structure. There are various ways to do this. The possibility  most often discussed in the literature is supposing the underlying blockmodel $\theta_{ab}$ stable over time, and assume that the nodes change cluster memberships over time according to some stochastic process, that is, suppose a specific time series structure for the indicators $Z_i(t)$, for example, a Markov process \cite{matias2017statistical, olivella2021adynamic}. A blockmodel varying in time is less frequently discussed, since this may make the model non-identifiable \cite{matias2017statistical} if at the same time the nodes are allowed to change cluster. However, \cite{jiang2020autoregressive} proposes an autoregressive network model with changepoints over time in the blockmodel, and \cite{pensky2019dynamic} discusses the theoretical properties of models with smoothly varying connectivity probabilities. Finally, to obtain interesting dynamics, one can also relax the conditional independence in the generation of the Bernoulli variables $A_{ij}$ over time. The simplest way of doing this is by introducing an autoregressive structure in the edge formation, making the value of $A_{ij}(t)$ directly depend on its previous measured value. This is done by imposing a first-order discrete autoregressive dependence in \cite{jiang2020autoregressive}, and by imposing a continuous-time Markovian process CAR(1) in \cite{ludkin2018dynamic}.

The above mentioned models estimate the network structure  assuming nodal clusters and a unique membership of each node at any time. However, in real data, nodes may belong to more than one cluster \cite{palla2005uncovering, palla2007quantifying}. Link communities, that is, when instead of nodes, edges are assumed to belong to one of a set of possible clusters in a data set, were originally proposed as a solution to this problem \cite{ahn2010link}, since thus every node can maintain links belonging to different communities. Dynamical link communities are discussed in, for example, \cite{meng2016incremental, nguyen2011adaptive} using one-by-one updates of an initial link community state of the network, but no statistical inference is drawn about the network and its parameters.

In our paper, we introduce the link blockmodel (LBM) based on an analogy with the SBM. We will impose, as with the SBM, a Bernoulli model for the state of the edges over time, but, relaxing the assumption of conditional independence, we will assume an (arbitrary-order) parametric autoregressive process to hold for the successive states of an edge (a link). Furthermore, we will suppose that the different edge clusters (the link communities) are characterised by their different parameter sets, similarly to the SBM, where the node clusters are distinguished by the (static) ``switch-on" probabilities comprised in the blockmodel $\theta_{ab}$. We present the model in the next Section \ref{sec:ll_analysis_corredge}.

\section{Likelihood Analysis of Correlated Edge Models}  \label{sec:ll_analysis_corredge}

\subsection{The logistic autoregressive model} \label{subsec:alarmll}

A class of basic models to deal with regression with a binary response variable is the generalised linear models (GLM). Its definition consists of the specification of the response distribution (the Bernoulli distribution), the linear predictor (comprising the influence of the covariates on the response), and the link function (the functional relationship between the linear predictor and the expected value of the response). We will use one such model to model the observed time series of an edge of a network, and combine these time series models into a blockmodel-type cluster structure based on the different dynamics of the time series.

Let $\bm{X}(t) = \{X_1(t), \ldots, X_D(t)\}^T$ denote a $D$-dimensional binary-valued vector with a multivariate Bernoulli dependence structure at time $t$. The collection of time series $\{\bm{X}(t)\}_{t=1,\ldots,T}$ satisfies an ALARM ({\it a} {\it l}ogistic {\it a}uto{\it r}egressive {\it m}odel \cite{agaskar2013alarm}) if its conditional probability distribution can be given as
\begin{align} \label{eq:general_alarm}
    &X_{i}(t) \, | \, \bm{X}(t-1), \ldots,  \bm{X}(t-K) \sim \mathrm{Ber}\left\{\mathrm{logit}^{-1}\left(\sum_{k = 1}^K \sum_{d = 1}^D b_{ikd} \, X_{d}(t-k) + c_i \right) \right\},    \\
    &b_{ikd},c_i \in \mathbb{R} \quad \mbox{for all } i,k,d, \nonumber
\end{align}
where $\mathrm{logit}^{-1}(x) = \exp(x) / [1+\exp(x)]$. The coefficients $ b_{ikd}$ represent the temporal dependence of $X_i$ on the previous values of the complete vector $\bm{X}(t-1), \ldots, \bm{X}(t-K)$, offering not only an autoregressive model for an edge, but also the possibility to model lagged cross-edge dependence. Depending on the value of the coefficients $ b_{ikd}$, a wide range of associations  can be modelled, including negatively correlated processes within the framework of one single model. The coefficients $c_i$ adjust the overall marginal probabilities of the component Bernoulli processes.

\begin{figure}[h]
    \centering
    \includegraphics[width=\textwidth]{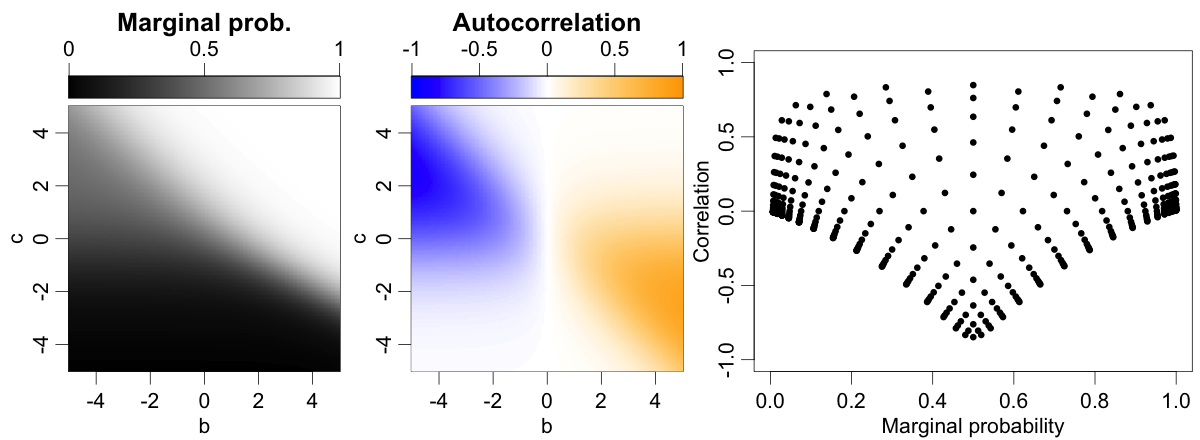}    
    \caption{Left and middle panels: the marginal probability and the  lag-1 autocorrelation of the ALARM(1) model \eqref{eq:alarm1} versus the model parameters. Horizontal axis: $b$, vertical axis: $c$. Right panel: The lag-1 autocorrelations versus the marginal probability for simulated  ALARM(1), using $b,c \in \{-5, 5\}$.}
    \label{fig:meancorr_vs_linpars_alarm1}
\end{figure}

The relationship between the parameters of the model \eqref{eq:general_alarm} and observable features such as autocorrelation and marginal probability of the realized process is not straightforward, especially not for higher autoregressive orders, but for an illustration, we show these for a range of parameter pairs for a first-order autoregressive ALARM model for $D = 1$, that is, a single binary time series:
\begin{align} \label{eq:alarm1}
    &X_t | X_{t-1} \sim \mathrm{Ber}\left\{\mathrm{logit}^{-1}\left(b \, X_{t-1} + c \right) \right\}, 
    \quad b,c \in \mathbb{R}. 
\end{align}
The middle panel of Figure~\ref{fig:meancorr_vs_linpars_alarm1} demonstrates that although negative values of the linear coefficient $b$ are associated with negatively correlated processes, and positive values with positively correlated ones, the relationship is not linear, and the value of the constant $c$ also has an influence on it. Moreover, maybe somewhat counterintuitively, both large negative and large positive $b$ values can produce near-zero autocorrelations when they are associated with large constants $c$ of the same sign. 

The right panel of Figure~\ref{fig:meancorr_vs_linpars_alarm1}, showing the lag-1 autocorrelation versus the marginal probability resulting from a range of parameter combinations, illustrates that while positively correlated binary time series can have any marginal probability, negatively correlated ones can have only much more restricted marginal probabilities as shown theoretically by \cite{teugels1990some,chaganty2006range}.

Statistical tests are needed to decide whether we should include autocorrelation into the model for a data set or not. One such test may be the comparison of the proportion of switched-on link states in the time series to an estimate of the edge probability based on the geometric distribution of the run lengths of the states. The two coincides  only under independence, since the run lengths will no longer have a geometric distribution if the time series is dependent. Another possibility is to check the validity of the geometric distribution for the run lengths, either by a simple quantile-quantile plot or by an (approximate) Kolmogorov-Smyrnov type distributional equivalence test. We will show an example of such a test for our data example in Section \ref{subsec:data_analysis}.

\subsection{The block-ALARM model} \label{subsec:balarmll}

Using the ALARM model, we can make the state of a link between two nodes depend directly on the state of the link at previous times. Link communities may then be assumed to follow distinctive temporal dependence models, with block-wise different parameters. 

Let us suppose we are dealing with a series of snapshots from an undirected network, observed at times $t_1,\ldots, t_n$, all on the same node set consisting of $N$ nodes. Let $A_{kj}(t_l)$ denote its (symmetric) adjacency matrix at time $t_l$. Define the (one-to-one) mapping $\sigma \, : \, \{(k,j) \, : \, k<j; \; k,j \in 1, \ldots, N \}  \mapsto \{1, \ldots, N(N-1)/2 \}$. Define the collection of random variables $\{X_{il}\}$ by the induced mapping $X_{il} = X_{\sigma(kj),l} = A_{kj}(t_l)$. Assume that each edge can belong to one of $G$ link communities, and let the variable $Z_i \in \{1, \ldots, G\}$ indicate the membership of edge  $X_{il}$ for all time indices $l$ (we assume that the membership of the edge does not vary over time). Our model, which we call block-ALARM (BALARM) model, can then be written as
\begin{flalign} 
    X_{il} \mid X_{i,l-1}, \ldots,  X_{i,l-K}, Z_i = g \sim \mathrm{Ber}\left\{\mathrm{logit}^{-1} ( \eta_{ilg} ) \right\}, \label{eq:alarm_modelform2} 
\end{flalign}
where $\eta_{ilg}$ is a linear predictor containing the characterisation of the system such as autoregressive terms, temporal patterns expressed by explicit functions of time (for instance a harmonic model), and covariates characterising the links. In the case when the model is supposed to contain only autoregressive terms, but no covariates or temporal patterns, its form is
\begin{flalign}  \label{eq:eta_ar}
\eta_{ilg} = \sum_{k = 1}^K b_{kg} x_{i,l-k} + c_g, 
\end{flalign}
where $b_{kg} \in \mathbb{R}$ represents the order $k$ autoregressive parameters in link community $g$, and $c_g \in \mathbb{R}$ determines the link probability value for link community $g$ when all the preceding $k$ time series values are zero. If the model is supposed to have a deterministic variation over time, such as in the presence of a typical daily pattern, this can be modified by adding terms containing time explicitly: 
\begin{flalign} \label{eq:eta_ar_with_t}
\eta_{ilg} = \sum_{d = 1}^D a_{dg} f_d(t_l) + \sum_{k = 1}^K b_{kg} x_{i,t-k} + c_g,
\end{flalign}
where $f_d(t)$ is an appropriate basis, for instance, harmonic functions in the case of a periodic temporal evolution.

Observing a collection of edges with unknown memberships, and supposing that an edge can be a member of a single cluster, we also assume a multinomial model for memberships: $Z_i \sim \mathrm{Multinom}(\pi_1, \ldots, \pi_G)$. Here $\pi_g$ is the probability of an edge to belong to cluster $g$. The complete-data likelihood of the model can then be written as \begin{flalign} 
L(\theta \; &; \; \{x_{il}\}, z_{i}) = \prod_{i = 1}^J \; \prod_{g = 1}^G \left\{ \pi_g \prod_{l = K+1}^n \left[ \frac{\exp(\eta_{ilg})}{1+\exp(\eta_{ilg})} \right]^{x_{il}} \left[ \frac{1}{1+\exp(\eta_{ilg})} \right]^{1-x_{il}} \right\}^{I(z_{i} = g)},
     \label{eq:alarm_sbm_completell} 
\end{flalign}
where the parameter $\theta$ represents all parameters $ \pi_g, a_{dg}, b_{kg}$ and $c_g$. The corresponding log-likelihood is
\begin{flalign} 
\ell(\theta \; &; \; \{x_{il}\}, z_{i}) 
= \sum_{i = 1}^J \sum_{g = 1}^G  I(z_{i} = g) \left\{ \log \pi_g + \sum_{l = K+1}^n \left[ x_{il} \eta_{ilg} - \log(1+e^{\eta_{ilg}}) \right] \right\},
     \label{eq:alarm_sbm_completelogl_stablemembership} 
\end{flalign}
with $\eta_{ilg}$ defined in Eq.~\eqref{eq:eta_ar_with_t}. This model can be fitted using the EM algorithm \cite{dempster1977emalgorithm}. Note that while the ALARM model as defined in \cite{agaskar2013alarm} allows us to include  cross-dependence between an edge and the lagged values of other edges, and this is, in principle, straightforward to do in this model too, we here do not assume such dependencies. A discussion of this follows in Section~\ref{subsubsec:crosscorr}.

\section{Simulation Studies and Data Analysis}

\subsection{Simulation Study} \label{subsec:simulations}

To explore the performance the model described above when applied to autocorrelated network data with a mixture of various edge probabilities ranging from moderately high to extremely low (similar to our data example), we simulated a series of BALARM models using the following ALARM(1) processes:
\begin{flalign*}
&\mathrm{Cluster \; A: \quad } b_1 = 2.89, \quad c_1 = -1, \quad p_1 = 0.67; \\
&\mathrm{Cluster \; B: \quad } b_2 = 4.48, \quad c_2 = -4, \quad p_2 = 0.045; \\
&\mathrm{Cluster \; C: \quad } b_3 = 5.43, \quad c_3 = -5, \quad p_3 = 0.016; \\
&\mathrm{Cluster \; D: \quad } b_4 = 6.42, \quad c_4 = -6, \quad p_4 = 0.006, 
\end{flalign*}
where $p_i$ denotes the stationary marginal probability of the resulting Markov chain. The lag-1 autocorrelation was set as 0.6 for all four processes. We created three two-component BALARM mixture models by combining Cluster A with each of the other three ALARM(1) models. 
This sequence represented a series of models in which all the edge processes had the same lag-1 autocorrelation, but they differed in their marginal link probabilities across a wide range, mimicking human interaction data in which link probabilities can range from very low to high, but where the persistence of edges is similarly high once switched-on. We generated $R = 200$ replicated data sets from each model, where each cluster contained 300 edge time series of a length of 1200 (totalling 600 edges per model). The length of the time series and the low edge probabilities were chosen to mimic the situation with our data set. We fitted each data set using the procedure described in Section \ref{subsec:balarmll}.

\begin{figure}[t]
    \centering
    \includegraphics[width=\textwidth]{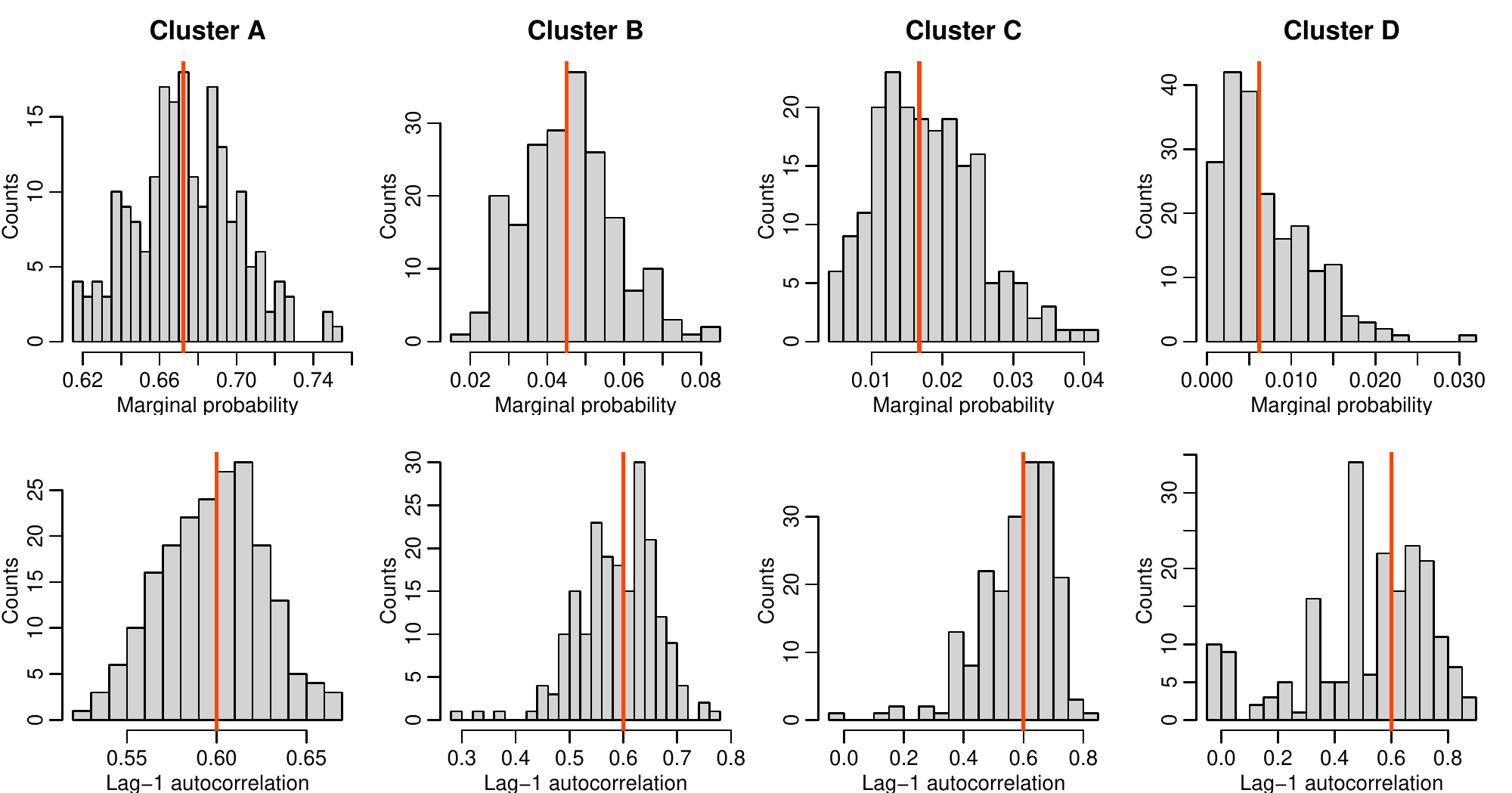}
    \caption{The distribution of the estimates of the marginal probability (upper row) and the lag-1 autocorrelation (lower row) for the four simulated clusters. The vertical red line indicates the true value of the parameter.}
    \label{fig:hist_margprobs}
\end{figure}

Estimates of the marginal probabilities and the lag-1 autocorrelations from the fits are shown in Figure~\ref{fig:hist_margprobs} (the histogram of cluster A, which occurred in all three models, is presented only once, although it was separately simulated and fitted for all three).  
Whereas the estimates for the clusters A and B appear to be reasonably good, the distribution of the estimated marginal probability for the two  low-probability clusters is asymmetric. This is expected as asymptotic normal theory of maximum likelihood estimates (which is approximated by the EM algorithm) breaks down for such low probabilities with the given time series length. Moreover, the estimate of the autocorrelation becomes unreliable, covering the whole $[0,1]$ interval, especially for cluster D with the lowest contact probability. This suggests that in data analysis with sparse contacts, which are quite typical in many applications of network analysis, we need to carefully consider the reliability of our estimates, and since asymptotic theory does not provide a sufficient quality of approximation, bootstrap methods are necessary.

\subsection{Data Analysis} \label{subsec:data_analysis}

Our data set contains the high-resolution dynamic network of social interactions in a hospital ward, taken with the aim to identify crucial spreaders in a hypothetical epidemics \cite{vanhems2013hospitaldata}. Social interactions between humans seem to be particularly in need to include correlations in their modeling, especially if observed in high temporal resolutions. Moreover, the strict daily schedules in a hospital imply daily varying contact probabilities between different groups in the hospital. The BALARM model, presented in Section \ref{subsec:balarmll},  imposes direct correlation between successive states of edges, and is adapted to provide a detailed model about the dynamics of the network over time, which can be particularly beneficial for modelling the unfolding of an epidemic.

\subsubsection{Data}

\begin{figure}
    \centering
    \includegraphics[width=\textwidth]{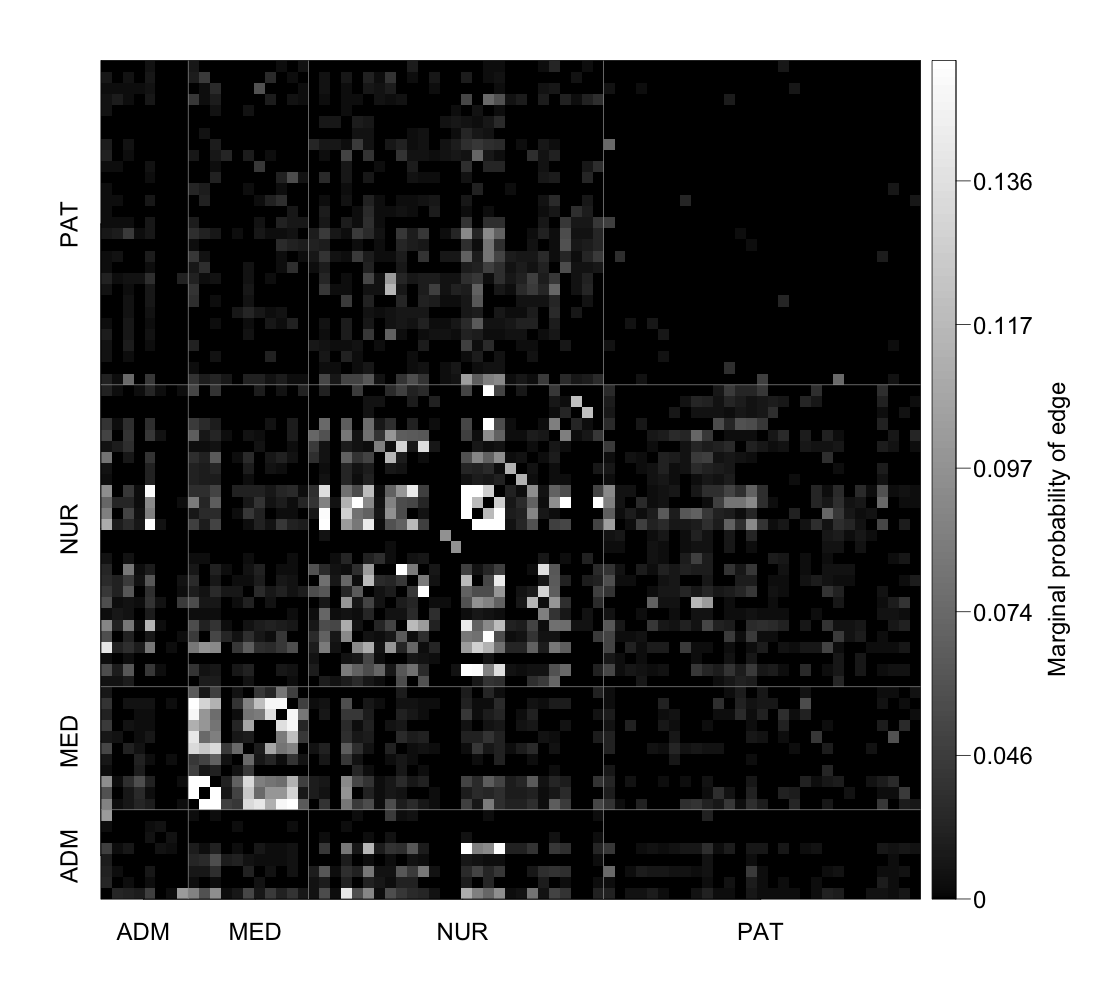}
    \caption{Link probabilities, averaged over time, between nodes of the hospital network. Shades of gray indicate the value of the probability, with black indicating 0 and white, 0.15. ADM: administrative staff, MED: medical staff, NUR: nursing staff, PAT: patients. }
    \label{fig:margprob_5min_1.5}
\end{figure}

Social interaction data over time from the geriatric short-stay ward of a university hospital in Lyon, France, was collected between Monday, December 6, 2010 at 1∶00 pm to Friday, December 10, 2010 at 2∶00 pm, using RFID (radio frequency identification) devices attached to 29 patients (coded PAT in what follows), 27 nurses (NUR), 11 medical doctors (MED) and 8 administrative staff (ADM), in total $N=75$ individuals \cite{vanhems2013hospitaldata}. These node categories will be termed ``status", following \cite{vanhems2013hospitaldata}. The RFID gives a contact signal if it is able to exchange radio signals with another RFID, which happens when their owners stand closer than about 1.5 meters from each other. This closeness was used as a proxy for a contact between two persons. Every 20 seconds, the presence or absence of  these contact signals during the preceding 20 second period were recorded between each pair of devices. For our use, we aggregated the data into 5 minute snapshots, by defining the adjacency value $A_{ij}(t_l) = 1$ at time $t_l$ between nodes $i,j \in \{1, \ldots, 75\}$ if there was at least one contact signal in the preceding 5 minutes between RFIDs $i$ and $j$, and  $A_{ij}(t_l) = 0$ if there was none. We used the time series $A_{ij}(t_l), l \in {1,\ldots, n}$ with $n=1159$ as our input data. The number of edges in the adjacency matrices is $N_e = N(N-1)/2 = 2775$.

\subsubsection{Model} \label{subsubsec:model}

The plot of the average edge probabilities arranged in an adjacency matrix format, shown in Figure~\ref{fig:margprob_5min_1.5}, indicates a block structure, which nevertheless does not fully coincide with the status of the nodes in the hospital, although a notable overlap exists in the case of doctors. This suggests that in this data, a hidden node cluster membership (related to but not identical with the hospital status) may explain at least part of the network structure within the framework of a time-varying stochastic blockmodel. However, a link community may, for example, provide the possibility of recognising sub-clusters of edges in different node communities with similar temporal patterns and correlations, or to scrutinize whether in this data example, status in the hospital fully determine the patterns of the interactions.

\begin{figure}[h]
    \centering
    \includegraphics[width=0.98\textwidth]{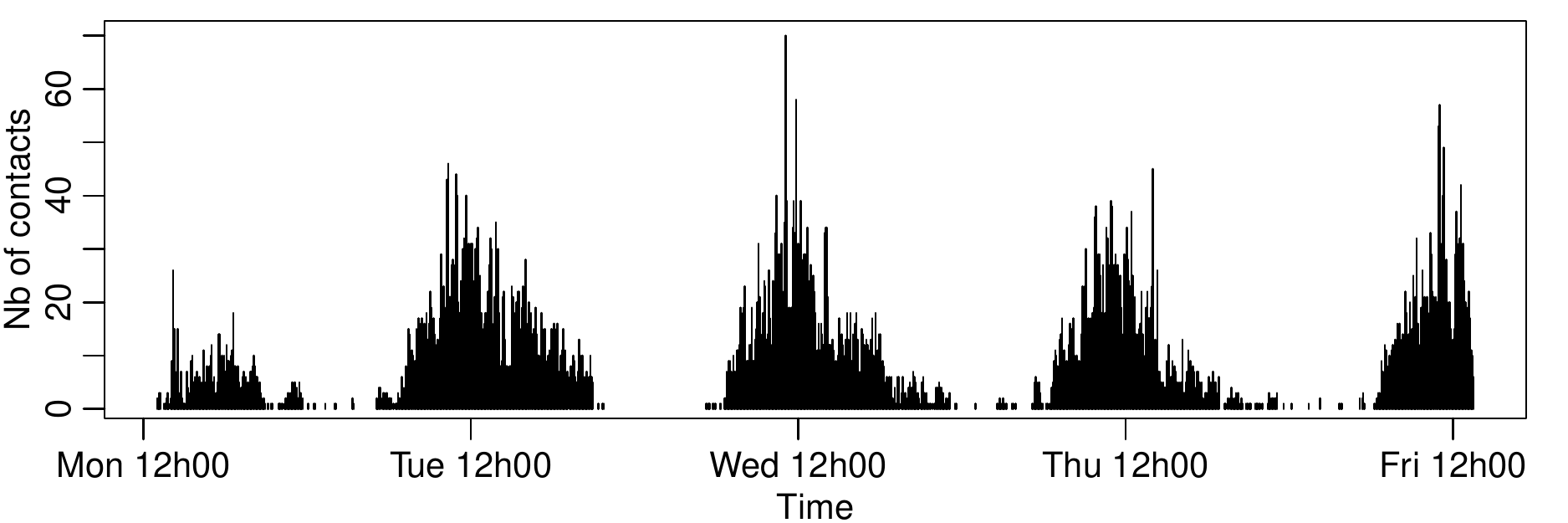}
    \caption{The total number of contacts  $\sum_{i\neq j} A_{ij}(t_l)$ in each 5-minute period within the observation span as a function of time.}
    \label{fig:ts_NbOfContacts_5min}
\end{figure}

\begin{figure}[h]
    \centering
    \includegraphics[width=\textwidth]{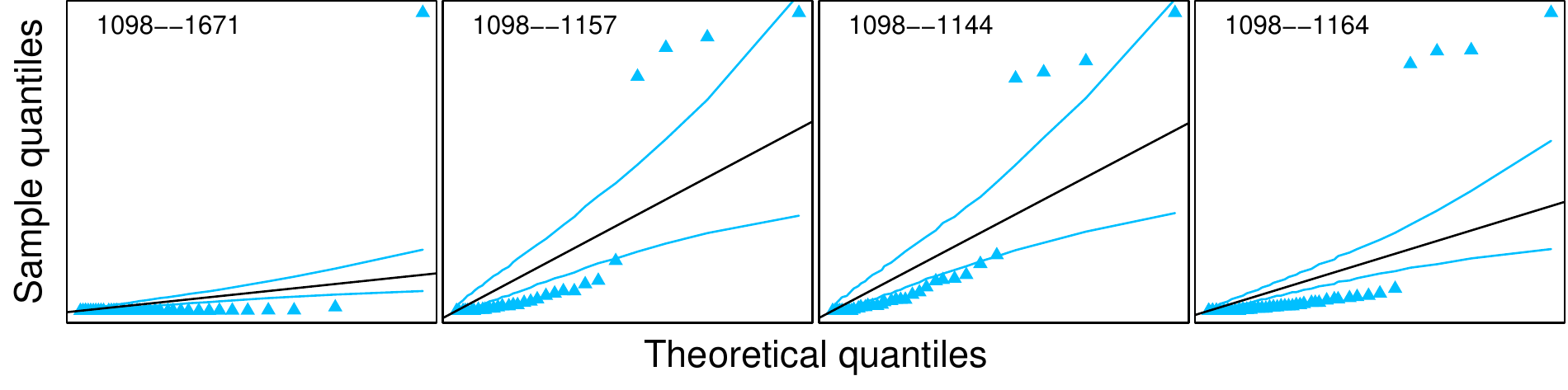}
    \caption{Geometric quantile-quantile plots for a few edges with link probability higher than 0.2. The black line has slope 1 and intersection 0, indicating the alignment of the expected positions of a sample from a geometric distribution.}
    \label{fig:geomQQ_5min}
\end{figure}

The plot of the number of contacts $\sum_{i\neq j} A_{ij}(t_l)$ versus $t_l$, shown in Fig.~\ref{fig:ts_NbOfContacts_5min}, suggests daily repeated patterns through the time span of the records, corresponding to the strict daily routine in a hospital. Since we are analysing human interaction data, which is typically autocorrelated, we should also test for the necessity of including an autoregressive term. For possibilities, we refer to Section~\ref{subsec:alarmll}.  For our data, we use geometric quantile-quantile plots, where we estimated the marginal probability of each time series as  $\hat p_{ij} = {\sum_{l = 1}^{T}{A_{ij}(t_l)}} / T$. We show these for some edges in Figure~\ref{fig:geomQQ_5min}, which indicates a discrepancy from the geometric distribution. However, in our case the visible presence of the periodically varying contact probabilities can also cause this. In our model, we will include an autoregressive term, and will test for its significance using bootstrap.

We therefore complemented the linear predictor of the model \eqref{eq:alarm_sbm_completelogl_stablemembership} with a $H$-order harmonic series with a period $P$ equal to a day ($P = 288$ in five-minute units):
\begin{flalign} 
\eta_{ilg} = \sum_{d = 1}^D a_{dg} f_d(t_l) + \sum_{k = 1}^K b_{kg} x_{i,t-k} + c_g,
     \label{eq:alarm_sbm_linpred_hospital} 
\end{flalign}
where $D = 2H$, $f_d(t) = \cos \left( 2\pi \left \lceil d/2 \right \rceil  P^{-1} t \right)$ for $d = 1,3, \ldots, 2H-1$ and $f_d(t) = \sin \left(2\pi (d/2) P^{-1} t \right)$ for $d = 2,4, \ldots, 2H$, and $\left \lceil . \right \rceil$ stands for the function ceiling. Moreover, as we  model the self-maintaining nature of human contacts, we suppose the autoregressive order to be $K = 1$.

The model was fitted using the EM algorithm \cite{dempster1977emalgorithm}, for a range of different choices for the number of link communities ($G = 2, \ldots, 9$) and harmonic order ($H = 2,3,4$). The best model was selected by the Bayes Information Criterion \cite{schwartz1978bic}, since BIC is a consistent selector of model complexity in clustering models and in linear modelling, and thus for our large data set, we can expect good performance. Moreover, according to  \cite{brewer2016therelative}, in cases like ours when the regression model is not expected to contain collinear variables, BIC slightly outperforms AIC and AIC$_c$ in terms of its power to select the correct model. We present the selected model fit in the next sections.

\subsubsection{Results}

\begin{figure}[h]
    \centering
    \includegraphics[width=0.8\textwidth]{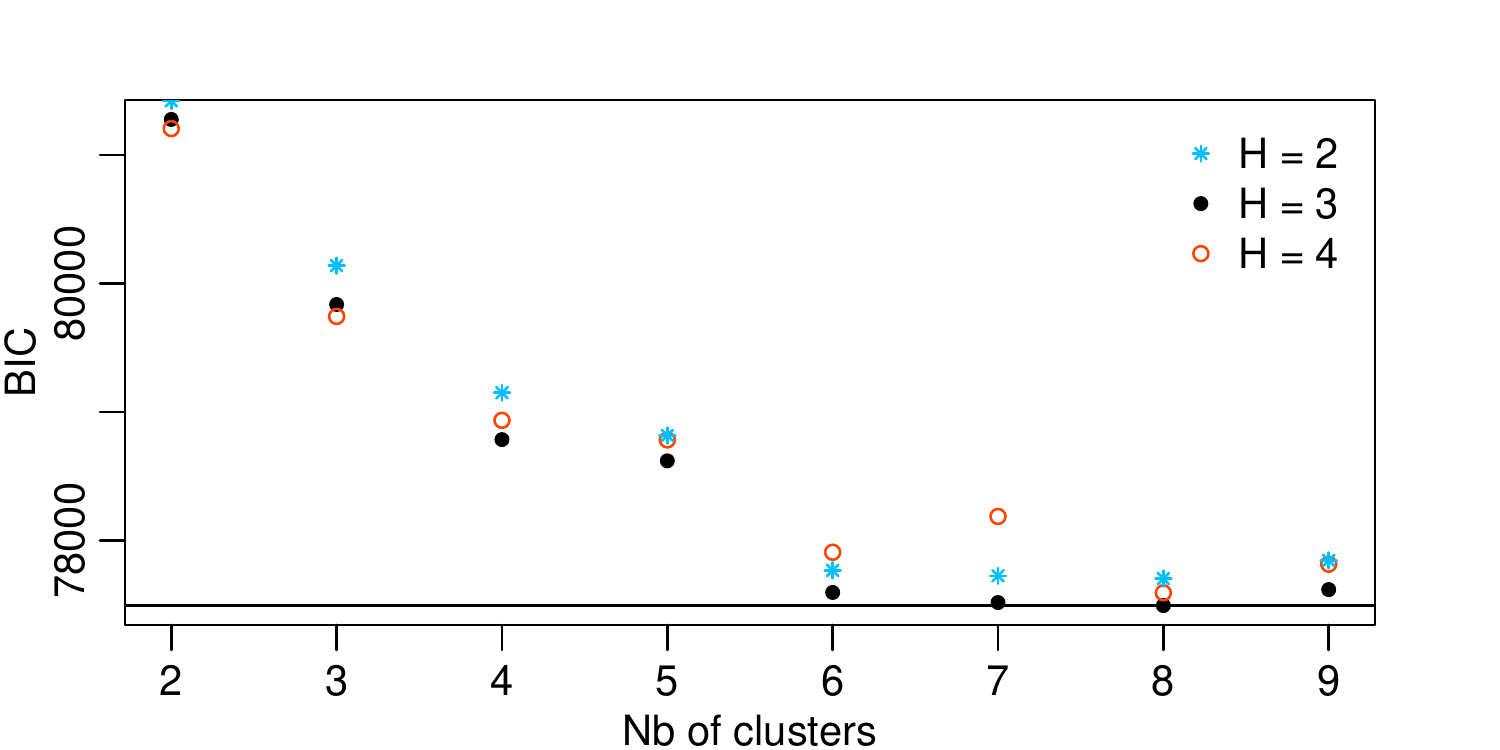}
    \caption{Bayes Information Criterion for the  BALARM(1) models fitted to the hospital data. The different harmonic orders $H$ are indicated with different colours and plotting symbols.}
    \label{fig:BIC}
\end{figure}

\afterpage{
\begin{figure}[]
    \centering
    \includegraphics[width=\textwidth]{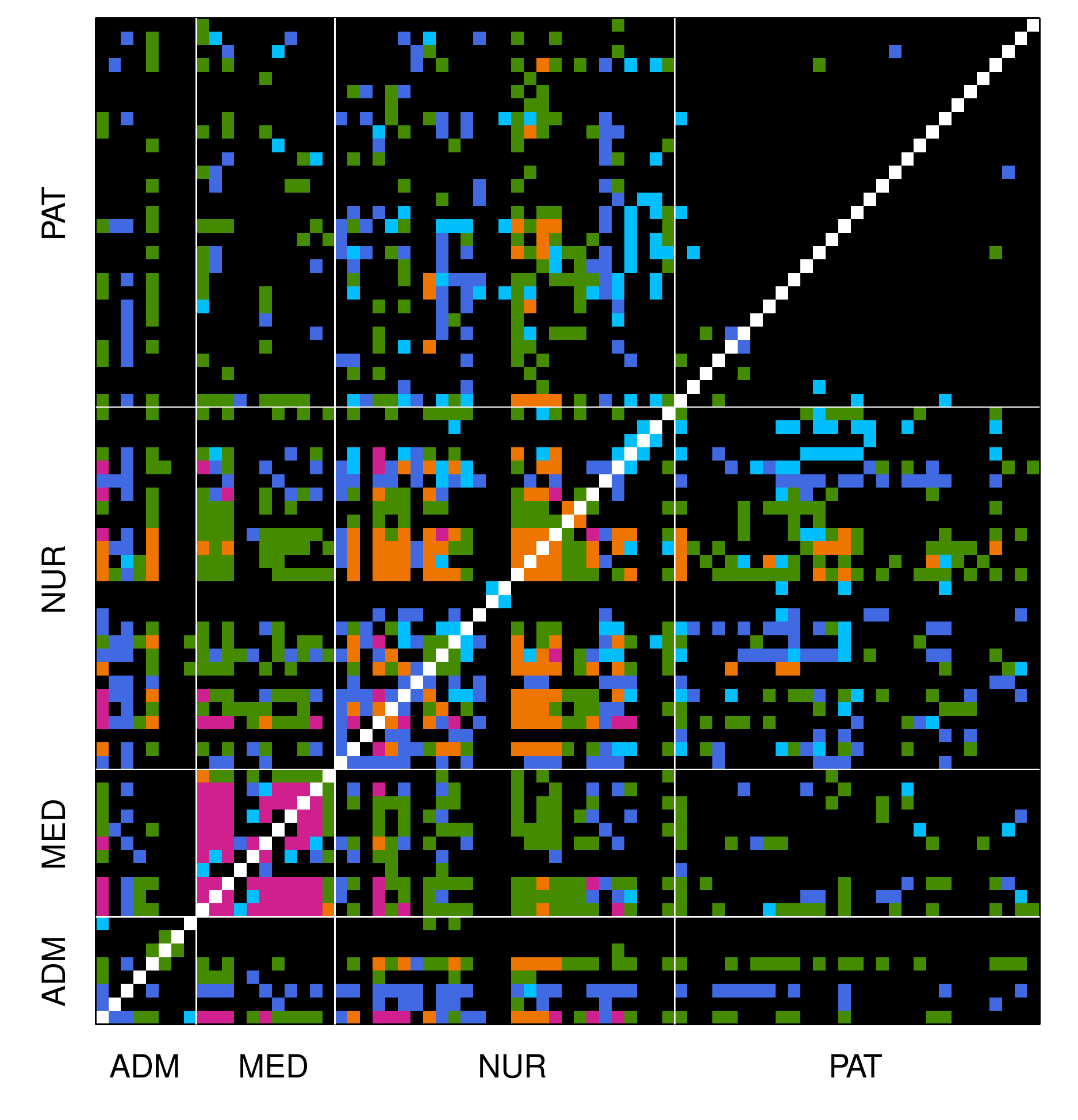}
    \caption{Cluster memberships of the edges (arranged in an adjacency matrix format) in the best model with 6 clusters and $H=3$ for the hospital data. The colors indicate the different link communities the edges belong to (black: LP, red: HPD, orange: HPM, green: MPM, light blue: MPD, dark blue: MPA; for the naming, see Table \ref{table:fitpars}). The arrangement of the nodes is the same as in Figure~\ref{fig:margprob_5min_1.5}.}
    \label{fig:clusterplot_6}
\end{figure}
\clearpage
}

Figure~\ref{fig:BIC} shows the resulting BIC values from the model fits. The overall best fit is the model with $H=3$ and $K=8$. However, the decrease in BIC values for $K > 6$ is small in comparison to the improvement on models with $K \leq 6$, and especially with the apparent best harmonic order $H=3$, the models are practically equivalent above $K=6$. Based on the principle of parsimony, we chose the model  with $H=3$ and $K=6$, as the model representing the best compromise between quality of description and simplicity. A summary of the basic parameters of the communities in this model and our notation for the clusters is given in Table~\ref{table:fitpars}.

\begin{figure}[h]
    \centering
    \includegraphics[width=\textwidth]{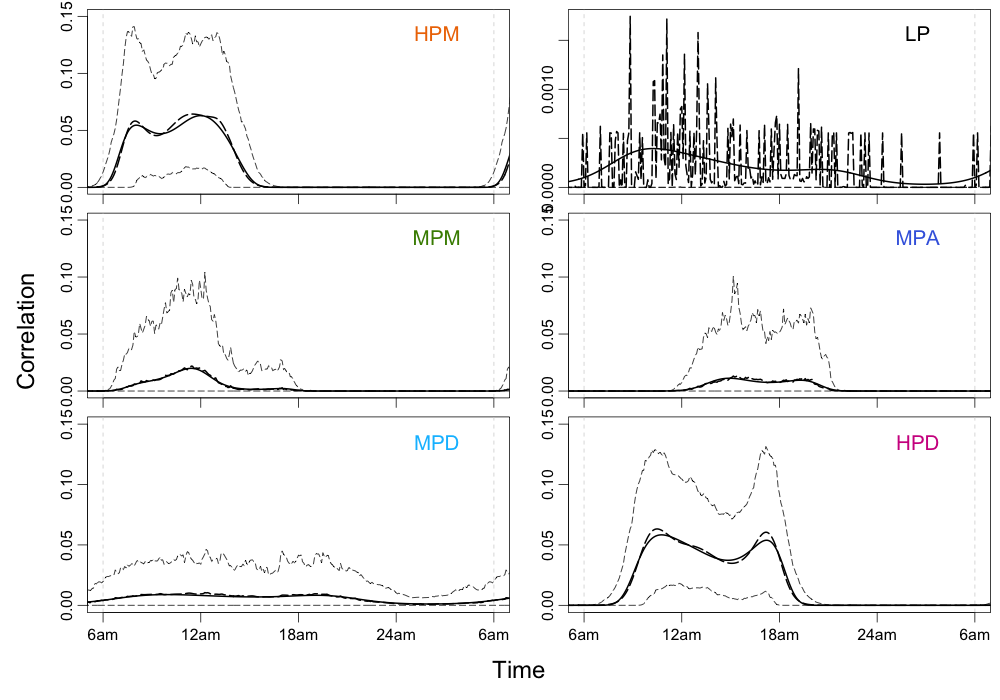}
    \caption{95\% pointwise bootstrap confidence intervals for the estimated daily variations of the link probabilities from the model with 6 components and 3 harmonic terms, for each identified cluster labelled according to Table \ref{table:fitpars}, and colour-coded according to Figure \ref{fig:clusterplot_6}. The thick solid line is the estimate on the real data. The heavy dashed line is the median of 500 bootstrap repetitions, the thin dashed lines represent the pointwise 0.025 and 0.975 quantiles. Note that while the five panels for the clusters HPM, MPM, MPD, MPA and HPD have common $y$-axis limits $[0,0.15]$, the upper right panel showing the LP cluster has different limits, $[0,0.0025]$.}
    \label{fig:margprob_boot}
\end{figure}

\begin{figure}[h]
    \centering
    \includegraphics[width=0.95\textwidth]{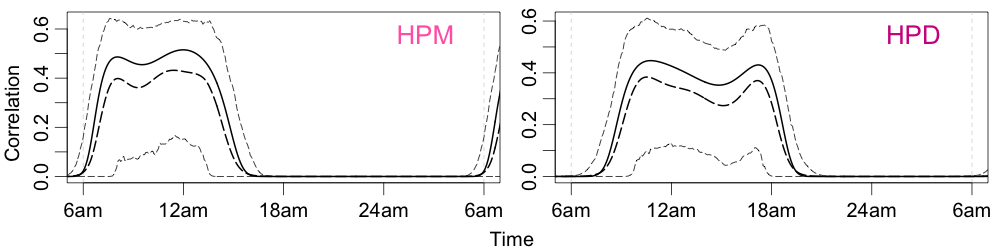}
    \caption{95\% pointwise bootstrap confidence intervals for the estimated daily variations of the lag-1 autocorrelation from the model with 6 components and 3 harmonic terms for the clusters with the highest link probabilities, for each identified cluster labelled according to Table \ref{table:fitpars}, and colour-coded according to Figure \ref{fig:clusterplot_6}. The line types correspond to the same quantities as presented in the caption of Figure~\ref{fig:margprob_boot}.}
    \label{fig:corr_boot}
\end{figure}

From the estimated model parameters, we can derive the average temporal variation of both the link probability and the lag-1 autocorrelation, and the most likely link community membership of each edge. The estimated memberships are shown in Figure~\ref{fig:clusterplot_6}, and the time-varying link probabilities and autocorrelations in Figures~\ref{fig:margprob_boot} and \ref{fig:corr_boot} (in heavy solid lines). The very low contact probabilities are reflecting the fact that in the data set, most (nearly 60\%) of the edges have no contacts at all through the whole observation period. These links are appropriately attributed to the LP cluster, for which, accordingly, the maximal contact probability in any 5 minute interval during a day is estimated at only 0.0004. Those links that do have at least one contact over this time have on average around 3 contacts during the data taking. The highest link forming probabilities belong to the HPM and HPD communities, still with a value of only about 0.06. With such low probabilities, we will resort to bootstrap for inference on the estimated daily patterns and autocorrelations.

We performed a parametric bootstrap analysis. We simulated 500 repetitions of the model using the estimated value of the parameters, and repeated the estimation on them. The initial values for the EM algorithm were fixed at their true value, to facilitate the identification of clusters. Finally, from the estimated bootstrap parameters, we reconstructed the temporal pattern of the marginal probabilities for all six clusters and the correlations for the two clusters with the highest link probability levels, for which we can hope for a realistic correlation estimation. The results are shown in Figures~\ref{fig:margprob_boot} and \ref{fig:corr_boot}.

\begin{table}
\caption{ Summary of the estimated link communities. The first two columns give the  name and a general characterisation of the six communities. The third column contains their  population as a percentage of the total number of possible links. The fourth column shows  their maximal switch-on probability during the day. \label{table:fitpars}}
\centering
\fbox{
\begin{tabularx}{0.85\textwidth}{*{4}{c|c|c|c}}
\em Cluster & \em Type & \em Percent & \em Max. link prob. \\
\hline
LP & {\it l}ow link {\it p}robability & 72\% & 0.0004 \\
MPM & {\it m}oderate link {\it p}rob., {\it m}orning  & 13\% & 0.02 \\
MPA & {\it m}oderate link {\it p}rob., {\it a}fternoon & 8\% & 0.011 \\
MPD & {\it m}oderate link {\it p}rob., {\it d}ay-long & 3\% & 0.008 \\
HPM & {\it h}igh link {\it p}rob., {\it m}orning & 3\% & 0.063 \\
HPD & {\it h}igh link {\it p}rob., {\it d}ay-long & 2\% & 0.058 \\
\end{tabularx}
}
\end{table}

Figure~\ref{fig:margprob_boot} shows that the 95\% pointwise confidence bands are quite wide, but broadly support the estimated daily patterns, such as with the two-peaked aspect of clusters HPM and HPD, the low-activity tail of MPM stretching into the early afternoon, and the day-long, moderate-level activity of MPD (for the definitions, please see Table \ref{table:fitpars}. The median of the bootstrap estimates matches very well the estimates on the real data, which indicates that the method estimates the contact probabilities in a reliable way. The broad uncertainty bands are not surprising, given the sparse contacts due to the extremely low contact forming probabilities together with a small degree of freedom. The lower limit of the bands, despite its appearance, does not include $p=0$, since the inverse logit transform does not allow for the probabilities to be precisely 0 or 1. However, they can get arbitrarily small. Longer observational time would be necessary to put a more stringent lower limit on the estimates. The spiky look of the confidence bands in panel LP is due to the combination of the extraordinarily low link probability (at most $0.0004$) of the cluster and the inverse logit transformation. 

It is not reasonable to calculate the lag-1 autocorrelations for the four low-probability clusters MPM, MPD, MPA and LP using time series of this length, as our simulations in Section~\ref{subsec:simulations} illustrated. Nevertheless, it can be calculated and estimated for the two clusters with the highest link probabilities, as shown in Figure~\ref{fig:corr_boot}. The daily pattern of the correlations are again supported by the bootstrap estimation, and are markedly bounded away from 0, indicating that it is indeed necessary to include an autoregressive term into the modelling of this data set. However, the correlations appear to be estimated with a negative bias. This suggests an even stronger correlation of human relations in reality than our model estimates, and underlines the importance to incorporate this autoregressive nature into modelling efforts.

\subsubsection{Interpretation}

The model fit offers a very detailed insight into the social network of a hospital ward. Several interesting conclusions can be drawn.

\paragraph{Relationship to hospital status.} The  patterns discovered in Figure~\ref{fig:clusterplot_6} are similar to the block patterns suggested by the time-averaged link probabilities in Figure~\ref{fig:margprob_5min_1.5}. It appears thus that the link clustering performed by the BALARM model is at least partly based on the time-averaged link probability of the different edges, and overlaps with what we would expect from a stochastic blockmodel which is based on the similarity of contact probabilities within blocks. However, the resulting link community model does not coincide fully with the clusters defined by status in the hospital. 

\paragraph{Discrimination of link communities based on different edge dynamics.} The model does not exclusively base its decision on the time-averaged contact probability of the edges. Most dynamic SBMs in the literature \cite{ludkin2018dynamic, matias2017statistical, matias2018asemiparametric, olivella2021adynamic, pensky2019dynamic} suppose that the edge probabilities of the different blocks are constant over time, and the dynamics of the networks are determined by the latent process of the nodes moving among the blocks. As opposed to this, the blocks in our model are discriminated based on their different time series characteristics, as our model definitions \eqref{eq:alarm_sbm_completelogl_stablemembership} and \eqref{eq:alarm_sbm_linpred_hospital} imply. Figure \ref{fig:margprob_boot} shows this very clearly: the daily variations of the contact probabilities of the six link communities are visibly different. For example, although the HPM and the HPD clusters have a qualitatively different two-peaked shape, the HPM cluster starts activity earlier than the HPD, at about 7 am when the HPD is still very close to 0 link probability confirmed by its bootstrap confidence bands. The activity of HPD cluster lasts longer than the HPM, being still highly active around 5pm when the HPM community has already ceased to be active.

\paragraph{Realistic picture of the dynamics.} The structures in Figure \ref{fig:clusterplot_6}, together with the dynamics in Figure \ref{fig:margprob_boot} capture many realistic details from the life of a hospital, which lends credibility to the model fit. 
\begin{itemize}
    \item We have identified two clusters with the most frequent contacts, HPM (orange in the figures) and HPD (purple). Figure \ref{fig:clusterplot_6} shows that one of them corresponds mostly to interactions between doctors, and the other is mostly associated to nurse-nurse interactions. It appears that as far as doctor-doctor interactions are concerned, they form their own near-exclusive block. Those doctors not following the same daily pattern in their interactions (a mostly black and a mostly green row in the MED-MED block in Figure \ref{fig:clusterplot_6}) may be a nutritionist and a physiotherapist who, according to the description of the data set in \cite{vanhems2013hospitaldata}, visited the ward occasionally, but were not present as regularly as the resident medical staff. They also have more sporadic interactions with the nursing staff than the other doctors, and less contact with patients and the administration too. Those doctors belonging to the main cluster HPD have not only very similar interaction patterns and link probabilities among themselves, but quite similar interaction patterns with nurses (those edges belonging mostly to the cluster MPM, green in the plots), and with patients too. Perhaps somewhat surprisingly, their edges with patients belong in majority  to the cluster LP (black in the plots) and to the cluster MPM, the two clusters with the lowest contact probabilities. We can also draw the conclusion that for the modelling of the interaction of doctors with everybody else, a 2-component SBM might be an adequate model. 
    \item Link structure within the block of nurses is far more complex. The NUR-NUR block is itself sub-divided into two large and at least one smaller block. (1) Some nurses interact with each other mostly in the morning (a mostly green and orange block along the diagonal, with HPM (orange) or MPM (green) contact probability patterns). Their interactions with patients also follow the morning patterns. (2) Another group interact within itself rather in the afternoon (predominantly dark blue block along the diagonal indicating the MPA cluster), though with more mixing of morning and afternoon patterns. Contacts with patients belong also to the MPA cluster.
    
    The interaction of these two groups of nurses mostly goes by the morning patterns. This probably reflects both a  division of the nurses into morning and afternoon shifts, and the existence of an overlap between the shifts, reasonable conclusion because information about the patients must be passed on some time.
   
    \item An anomalous group of nurses can be found as mostly black rows in the middle and at the top of the NUR-NUR block in Figure~\ref{fig:clusterplot_6}. Almost all their contacts with other staff members belong to the LP (black) cluster. With patients however, many of their links belong to the cluster not mentioned so far, the MPD cluster (light blue in the plots). More complex models, such as the BIC-best 3-harmonic, 8-cluster model or the 4-harmonic, 6-cluster model, identify the NUR-PAT links of these nurses as a separate link cluster with a specific daily pattern consisting of a morning and an evening burst of activity. These bursts in these more complex models reach the highest link probabilities earlier in the morning and later in the evening than any other links. Based on this, we might guess that our model has identified the nurse-aides, those who have fewer medical tasks than the regular nurses or none at all, but care about the patients' basic needs such as getting dressed, washed or fed, possibly before or after the medical needs of the patients are satisfied during the workday.
    \item Doctors and nurses interact mostly in the morning, according to the MPM (green) link cluster pattern, with on average lower link probability than doctors have with other doctors, and a definitely different temporal pattern. Some links however belong to the MPA (dark blue) afternoon pattern. These two moderate link probability patterns make up those MED-PAT interactions too which do not belong to the low-probability LP (black) cluster.
    \item Patients have almost no contact with each other. They also have on average much fewer contacts with anybody else than the others with each other. They have the most frequent contacts with the nurses, but even that does not reach the average level of interactions between nurses and doctors. This, although striking for a first sight, is perhaps expected. Patients in a geriatric short-stay ward may be seriously ill, affected by neurodegenerative diseases, and generally not in the mood of making contacts beyond the necessary (visitors were not tagged with an RFID (radio frequency identification device), and were not followed in the experiment).
\end{itemize}

This data set was also analysed by \cite{jiang2020autoregressive}. In that study, the researchers used a rougher aggregation of the data than we did, taking $A_{ij} = 1$ if there was at least one contact between nodes $i$ and $j$ during the day. This was needed since their model is stationary by construction, but the data at aggregations finer than one day are not stationary but have strong cyclostationary features. They find no evidence for the existence of nodal communities, and no evidence of significant autoregressivity. It is not surprising that our findings differ from those of \cite{jiang2020autoregressive}, as the absence of autoregressivity at the timescale of a day is plausible in the data. We expect human contacts to be strongly correlated on short timescales ($\sim$ minutes), but much less so over days, if not in specific circumstances. As to the nodal clustering, our conclusion that there may be an approximate SBM-like block structure at least for a MED--(everyone else) division relies strongly on the estimated sub-daily contact probability patterns, and thus may remain undetectable with long aggregation times.

\subsubsection{Cross-correlations between edges} \label{subsubsec:crosscorr}

\begin{figure}[h]
    \centering
    \includegraphics[width=\textwidth]{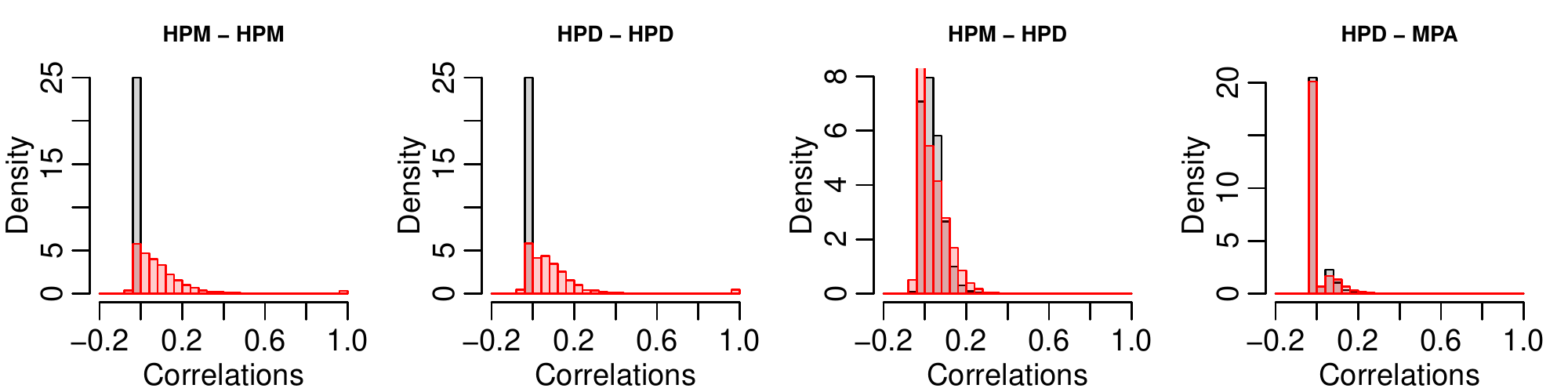}
    \caption{Histogram of cross-correlations between edge time series belonging to different link communities from the real data (red) and from simulations from the fitted model (grey).}
    \label{fig:crosscorr_lag0_hospital_sim}
\end{figure}

The question can be asked whether we need to include between-edge (possibly lagged) cross-correlations as well as the temporal autocorrelations. In the presence of periodically varying link probabilities of the clusters, apparent correlations may be found simply due to the similarities in the temporal patterns of some edges: contacts may be observed simultaneously simply because of their higher probability at some times at some lags, giving rise to spurious cross-correlations. To check for this, we simulated 5000 time series of a length equal to the observed data, independently from each of the clusters of our model, and computed cross-correlations between them. 

The main results are presented in Figure~\ref{fig:crosscorr_lag0_hospital_sim}. The only strong difference between the simulations with no cross-correlations (in grey) and the real data (red) was between two HPM or two HPD edges, as shown in the left two panels of Figure~\ref{fig:crosscorr_lag0_hospital_sim}. This suggests that cross-correlation might exist between such link types. However, it is also possible that some insufficiently modelled temporal patterns give rise to these apparent excess correlations. 

We found only small discrepancies for any other edge combination, of which two are shown in the right-hand panels (HPM-HPD and HPD-MPA). The existence of cross-correlations is, of course, not excluded, our finding may just mean that similarly to the autocorrelation, for such low contact probabilities the estimation of cross-correlation needs more observations.

\section{Discussion}

In our study, we proposed an approach combining binary-valued time series analysis with mixture modelling, in order to model the dynamics of networks describing human interactions. The nature of the data, which was collected in a hospital ward during four workdays with a high temporal resolution, raised an important modelling questions. Namely, how to account for the likely strong temporal autocorrelation and repetitive time-varying patterns present in the data, and in human interactions in general? Conditional independence and thus the connecting  graphon framework may not be able to adequately model the strong and direct autoregressivity of human contacts: we do not decide randomly and independently at each moment whether we continue a conversation. This leads us out of the most often used family of models, the SBM \cite{matias2017statistical,xu2014dynamic,olivella2021adynamic,pensky2019dynamic}, which are based on conditional independence. 
We proposed our block link community model (which we call BALARM) using the generalised linear model framework in reply to this question, and in order to explore its potential in real-life situations, we applied it to the hospital data.

Our BALARM modelled the time series of the elements of the adjacency matrix of the network as an autoregressive binary time series, assuming that each of these edges belonged to a single link cluster. The time series model for each edge included deterministic, time-varying linear components with a period of a day to account for the workday patterns at the hospital, and a linear  autoregressive term, supposing that conditioned on the immediate past, the contacts' existence does not depend on the more distant past (a reasonable assumption for these contacts focusing mostly on current events and problems in the hospital), but it does depend on the immediate past. The logit transformation linked this linear predictor to the instantaneous probability of the edge to become switched-on. We further assumed that there is a finite number of link communities (clusters) the edges can belong to, and that within these clusters, the parameters of the time series are constant. Furthermore, we supposed that the model is identifiable in the sense that at least one parameter differs for two different link clusters.

We fitted the model using the EM algorithm for a range of model complexities, and selected the fit realizing the best trade-off between simplicity and richness of interpretation according to the BIC. We obtained inference about the quantities of interest such as the daily contact probability patterns of the different communities and the autocorrelation over time produced by the fit by bootstrap. The results gave an unprecedentedly detailed insight into the time variations of the contacts of a social network. We could distinguish six clusters of different typical link probability variations over the day. These daily patterns could be put into realistic correspondence with the normal working day in a hospital, and also with the status of the persons in the hospital that formed them. We observed that although much of the link community structure found was clearly related to status in the hospital (namely, medical or administrative staff, nurse or patient), no perfect correspondence with an SBM could be found.

Our approach differs in several aspects from the currently available models \cite{jiang2020autoregressive, ludkin2018dynamic} that explicitly take into account the autocorrelated nature of the links. The most important of these is that we have deliberately stepped out of the usually considered framework of node clustering, and instead, considered edges as the basic units of modelling. Both of \cite{jiang2020autoregressive} and \cite{ludkin2018dynamic}, although they include autoregressivity directly into the network's temporal dynamics, is based on node clustering, although for technical reasons \cite{ludkin2018dynamic} does effectively use link clusters within their implementation.  Our model thus provides an alternative to these models for social situations where the nature of the links themselves is the subject of investigation. 

Another important difference is that the time series model at the core of our model is the classical generalised linear model framework \cite{mccullaghnelder1989generalized}, which has not yet been used in the modelling of dynamical networks, and which confers several significant advantages to our model. First, it offers the possibility to fit both negatively and positively correlated networks within one single model. This has a strong relevance in practice, when trying to find link communities in real data. If it is possible that the data contain both positively and negatively correlated time series, most models currently in use need to set up two formally different models, which causes difficulties such as obtaining correct inference including uncertainty arising from the decision about positivity or negativity of correlation in the model. Our model  accommodates naturally both possibilities within a single modelling framework. Another advantage provided by our framework is the flexibility to specify the temporal evolution of the network. Higher-order autoregressive terms and further relevant covariates, such as the harmonic terms in our model, can be straightforwardly added to the linear predictor of the model, and thus the detailed analysis of both stochastic and deterministic components of the dynamics becomes possible. This flexibility allowed us to decipher a realistic image of the life of the hospital ward under investigation from the network data, and to detect communities and distinguish their daily patterns, where the SBM-based model of \cite{jiang2020autoregressive} hit the problem of non-stationarity.

In principle, our model allows also for the inclusion of cross-correlation terms into the mode, leading effectively to a model with similarities to spatio-temporal models. However, care needs to be taken then how to select the basis in which we express the model's various temporal and spatial characteristics. Identifiability issues can arise due to confounding between the cross-correlation terms and the deterministic time variations due to similarities in time patterns. Another possible generalisation is the adaptation of mixed membership models for link communities, which may be important and interesting for applications where edges cannot be supposed to belong to one single link community. An example of this may be the contacts between two people whose contacts are at times "collaborator" type and at other times "friendship" type. This needs to model edges which change membership over time. Mixed membership models could be the correct way to model such social networks, however in such models, identifiability issues must be considered carefully.

We believe that our model provides a level of insight into the link community evolution that has not been previously reached yet. Moreover, it offers a model-based, controllable simplification to compress useful information of observed complex networks, which can be used as a building block to gain insight into processes on the networks, and is amenable to statistical inference. Such realistic model fits can serve, for example, as background social networks in in-depth investigations of the spread of an infectious disease within a social unit, providing a detailed insight into the evolution of the disease in the  community, and helping identify the most efficient intervention points. Their practical use, we hope, will be a strong spur in the future to develop both more easy-to-apply, useful, realistic models and the theory behind dynamic link community models.

\end{document}